\documentclass[12pt]{article}\usepackage[hyperfootnotes=false]{hyperref}
\usepackage{epsfig}
\usepackage{float}
\usepackage{empheq}
\usepackage{bbold}

\usepackage[utf8]{inputenc}
\usepackage{amsmath}

\usepackage{caption}

\usepackage{amsmath}
\usepackage{amssymb}
\usepackage{graphicx}
\newlength{\abstractwidth}
\renewcommand{\thefootnote}{\fnsymbol{footnote}}
\renewcommand{\thanks}[1]{\footnote{#1}} 
\newcommand{\starttext}{
\setcounter{footnote}{0}
\renewcommand{\thefootnote}{\arabic{footnote}}}

\newcommand{\be}{\begin{equation}}
\newcommand{\bea}{\begin{eqnarray}}
\newcommand{\eea}{\end{eqnarray}}
\newcommand{\beq}{\begin{equation}}
\newcommand{\ee}{\end{equation}}

\newcommand*\widefbox[1]{\fbox{\hspace{2em}#1\hspace{2em}}}

\def\eq{&=&}

\def\simleq{\; \raise0.3ex\hbox{$<$\kern-0.75em
\raise-1.1ex\hbox{$\sim$}}\; }
\def\simgeq{\; \raise0.3ex\hbox{$>$\kern-0.75em
\raise-1.1ex\hbox{$\sim$}}\; }

\def\bi{\begin{itemize}}
\def\ei{\end{itemize}}

\def\t{\tau}

\def\bsub{ \begin{subequations}
\begin{empheq}[box=\widefbox]{align}  }
\def\esub{ \end{empheq}
\end{subequations}}

\def\1{\(  \mathbb{1} \)}

 \def\lf{\left(}
    \def\rg{\right)}

  \def\bn{\bigskip \noindent}

 \def\bm{\begin{bmatrix}}
 \def\em{\end{bmatrix}}

\makeatletter
\g@addto@macro\normalsize{%
  \setlength\abovedisplayskip{10pt}
  \setlength\belowdisplayskip{20pt}
  \setlength\abovedisplayshortskip{10pt}
  \setlength\belowdisplayshortskip{20pt}
}
\makeatother

\usepackage{color}

\begin{document}


  
\begin{titlepage}

\rightline{}
\bigskip
\bigskip\bigskip\bigskip\bigskip
\bigskip

\centerline{\Large \bf {  Three Impossible Theories         }} 

\bn

\bigskip
\begin{center}
\bf      Leonard Susskind  \rm

\bigskip
Stanford Institute for Theoretical Physics and Department of Physics, \\
Stanford University,
Stanford, CA 94305-4060, USA \\

and

Google, Mountain View, CA

\end{center}

\bn

\begin{abstract}

I will begin by conjecturing a cosmological generalization of black hole complementarity (also known as the central dogma). I will then discuss three theories and argue that they are inconsistent with second law of thermodynamics if the  cosmological version of the dogma is correct.   The three theories are: the big rip; cyclic cosmology; and the Farhi-Guth-Guven mechanism for creating inflating universes behind black hole horizons.

\end{abstract}

\end{titlepage}

\starttext \baselineskip=17.63pt \setcounter{footnote}{0}

\Large

\tableofcontents


\section{The Central Dogma}
My arguments in this note are based on a  ``central dogma" which I have borrowed and  adapted from a paper by Almheiri, Hartman, Maldacena, Shaghoulian and Tajdini   \cite{Almheiri:2020cfm}. In the case of \cite{Almheiri:2020cfm} 	the context was black holes and the statement of the central dogma is this\footnote{The central dogma is a restatement of postulate 1 of black hole complementarity \cite{Susskind:1993if}: ``The process of formation and evaporation of a black hole, as viewed
by a distant observer, can be described entirely within the context of standard
quantum theory. In particular, there exists a unitary S-matrix which describes
the evolution from infalling matter to outgoing Hawking-like radiation." }:

\bn
\it As seen from the outside, a black hole can be described in terms of a quantum system
with ${\rm {Area}} /(4G_N )$ degrees of freedom, which evolves unitarily under time evolution.\rm

\bn
The central dogma of this paper is analogous but applies to cosmological space-times with 
horizons---specifically to de-Sitter-like geometries:

\bn
\it As seen from a causal patch a cosmological space-time can be described in terms of an isolated quantum system
with\footnote{In this case the area refers to the cosmological horizon} ${\rm {Area}}/(4G_N )$ degrees of freedom, which evolves unitarily under time evolution.\rm

\bn
In particular I assume that an observer in a causal patch sees a world of finite entropy satisfying the second law of thermodynamics.

The terminology in the cosmological version of the dogma requires some explanation. First,  ${\rm{{Area}}}$ refers to the area of the cosmological  horizon. For de Sitter space with metric,
\be 
ds^2 =-\lf 1-\frac{r^2}{R^2}  \rg dt^2 + \lf 1-\frac{r^2}{R^2}  \rg^{-1}dr^2 +r^2 d\Omega^2
\ee
the area of the horizon is,
\be 
{\rm{{Area}}} = 4\pi R^2.
\ee

Second, by an isolated quantum system I  mean a system dynamically uncoupled to any other system. Dynamically uncoupled does not mean unentangled. For example in the case of the two-sided eternal black hole \cite{Maldacena:2001kr}  in anti de-Sitter space, the two sides are uncoupled by the Hamiltonian but are entangled in the thermofield double state. Being dynamically uncoupled does imply that  the entanglement entropy of the two sides is constant.

The black hole version of the dogma is, with good reasons, widely accepted. Less  is known about cosmological horizons, so we should consider the cosmological version to be a conjecture.
One important consequence of this conjecture is that the second law of thermodynamics must be respected by all observations within a causal patch. (The entropy that the second law refers to includes the entropy of matter as well as the horizon entropy.)

Naively the second law means the entropy never decreases. In fact since dissipation (friction) is ubiquitous and no process is perfectly  adiabatic, the second law can be strengthened: entropy always increases.

Unless it doesn't: entropy can decrease. What the second law really says is that the probability for entropy to not increase is extremely small. We can quantify it as follows:
The probability for a process to take place in which an isolated system goes from state $A$ to state $B$, with the entropy of $B$ being smaller than the entropy of $A$, is of order,
\be 
P_{A\to B} \approx e^{-(S_A -S_B)}.
\label{P=expDS}
\ee 

As an example we might consider a container with $N=10^{24}$ uniformly distributed gas molecules (state $A$). What is the probability that a gap forms in which ten percent of the volume becomes empty (state $B$)? In that case,
\bea 
S_A -S_B \eq N\log{V} - N\log{(.9V)} \cr \cr
&\approx& .1N 
= 10^{23},
\eea
and
\be 
P_{A\to B} \approx e^{-10^{23}}.
\ee 
That's a small number. For practical purposes we take it to be zero and say that the second law forbids such entropy-decreasing processes.

\bn

Another consequence of the second law, taken together with the ubiquity of dissipation, is that an isolated finite system will tend to thermal equilibrium, i.e., the state of maximum entropy. This implies that eternal periodic macroscopic oscillations are impossible. For example a mass suspended by a spring will eventually stop oscillating because the mechanical energy of oscillation will thermalize and heat the spring. This is often stated as the impossibility of a perpetual motion machine of the third kind.

\section{The Big Rip}

It is claimed that current observation favors an equation of state with $w<-1.$ The implications of this include that statement that after matter and radiation are diluted away the expansion due to dark energy will feature an increasing Hubble parameter. In other words in the future cosmologists will measure,
\be 
\frac{dH}{dt} >0.
\ee
This behavior is sometimes called ``the big-rip" \cite{Caldwell:2003vq}.  The problem  is that it also means the horizon radius and area will eventually start to decrease. Assuming the usual connection between area and entropy, the big-rip scenario requires the entropy of an observer's causal patch to decrease,
\be 
\frac{dS}{dt} <0,
\ee
a clear violation of the second law.

One way out of this conclusion would be if there is another system coupled to the causal patch which drains off entropy. That may or may not be possible but it does violate the cosmological central dogma which assumes that a causal patch is an isolated system.

\section{The Cyclic Universe}

The cyclic universe postulates an eternally bouncing scale factor. More specifically in the flat slicing of the universe,
\be 
ds^2 = -d\t^2 +a(\t) dx^idx^i
\ee
the scale factor behaves like
 \be
 a(\t) = e^{H\t + f(\t)} 
 \label{oscillations}
 \ee
 where $f(\t)$ is periodic and averages to zero over an oscillation. If $f(\t)$ is large enough the geometry describes a periodic series of bounces or a so-called ``cyclic universe" \cite{Steinhardt:2002kw}.
 
 Oscillations of this type might be expected to produce entropy and thereby heat the universe. However, over each cycle the universe grows by a factor $e^{H\Delta t}$ which, it is claimed, would dilute the entropy and keep the entropy density from growing. 
 
\begin{figure}[H]
\begin{center}
\includegraphics[scale=.3]{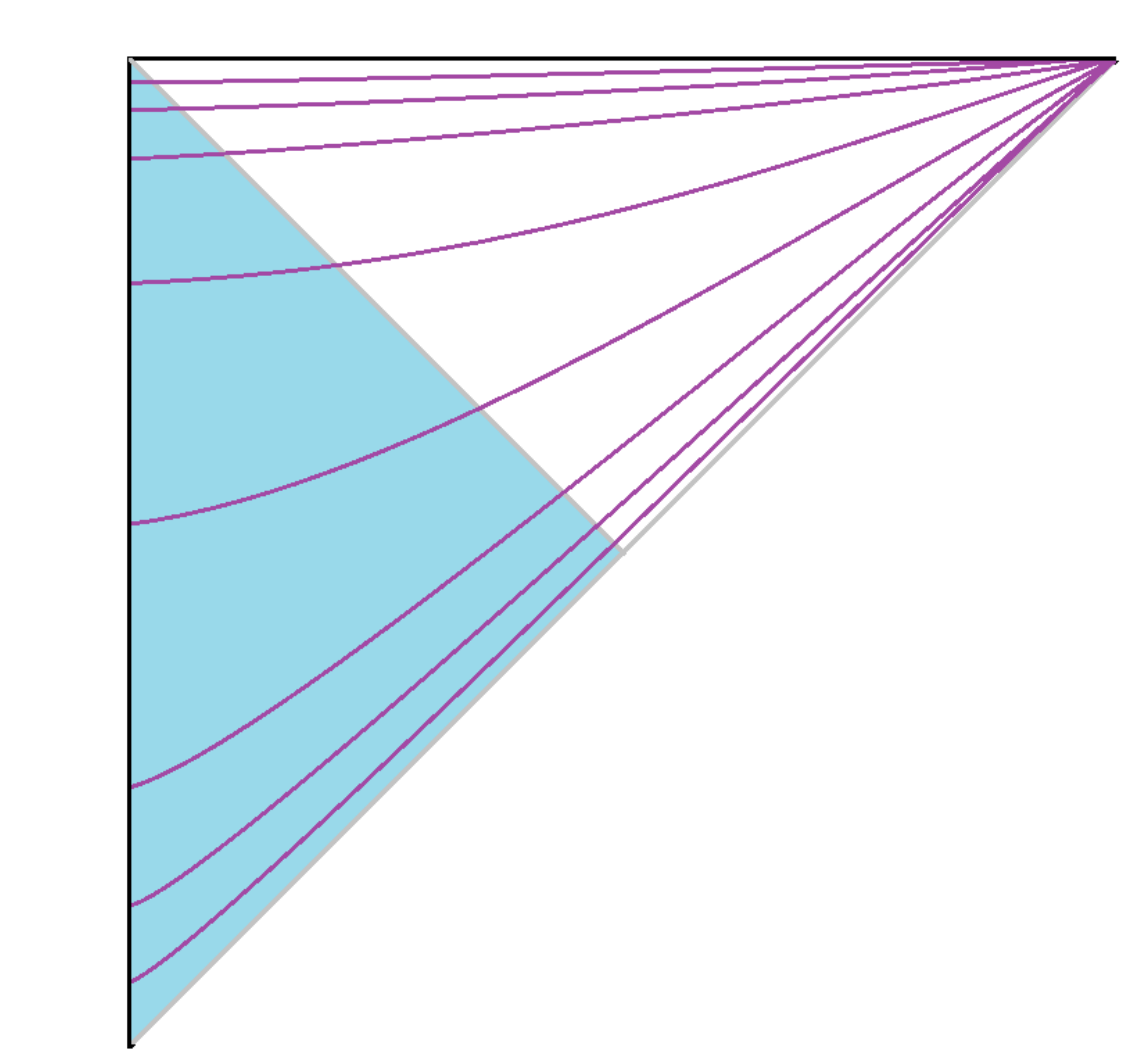}
\caption{Penrose diagram for a cyclic universe or a de Sitter space in flat slicing. The curved purple lines are flat time slices on which the bounces occur. The blue region is a causal patch. }
\label{halfDS}
\end{center}
\end{figure}

When averaged over a cycle the universe indeed grows exponentially according to \ref{oscillations}, just as it would in pure de Sitter space. In fact the Penrose diagram for the geometry is the same as for the flat slicing of de Sitter space and is shown in figure \ref{halfDS}.

The curved space-like lines are time slices but they can also be used to illustrate the time-dependent bouncing behavior; for example they can represent the bounces themselves, i.e., the  minima of $f(\t)$. 
 
\begin{figure}[H]
\begin{center}
\includegraphics[scale=.3]{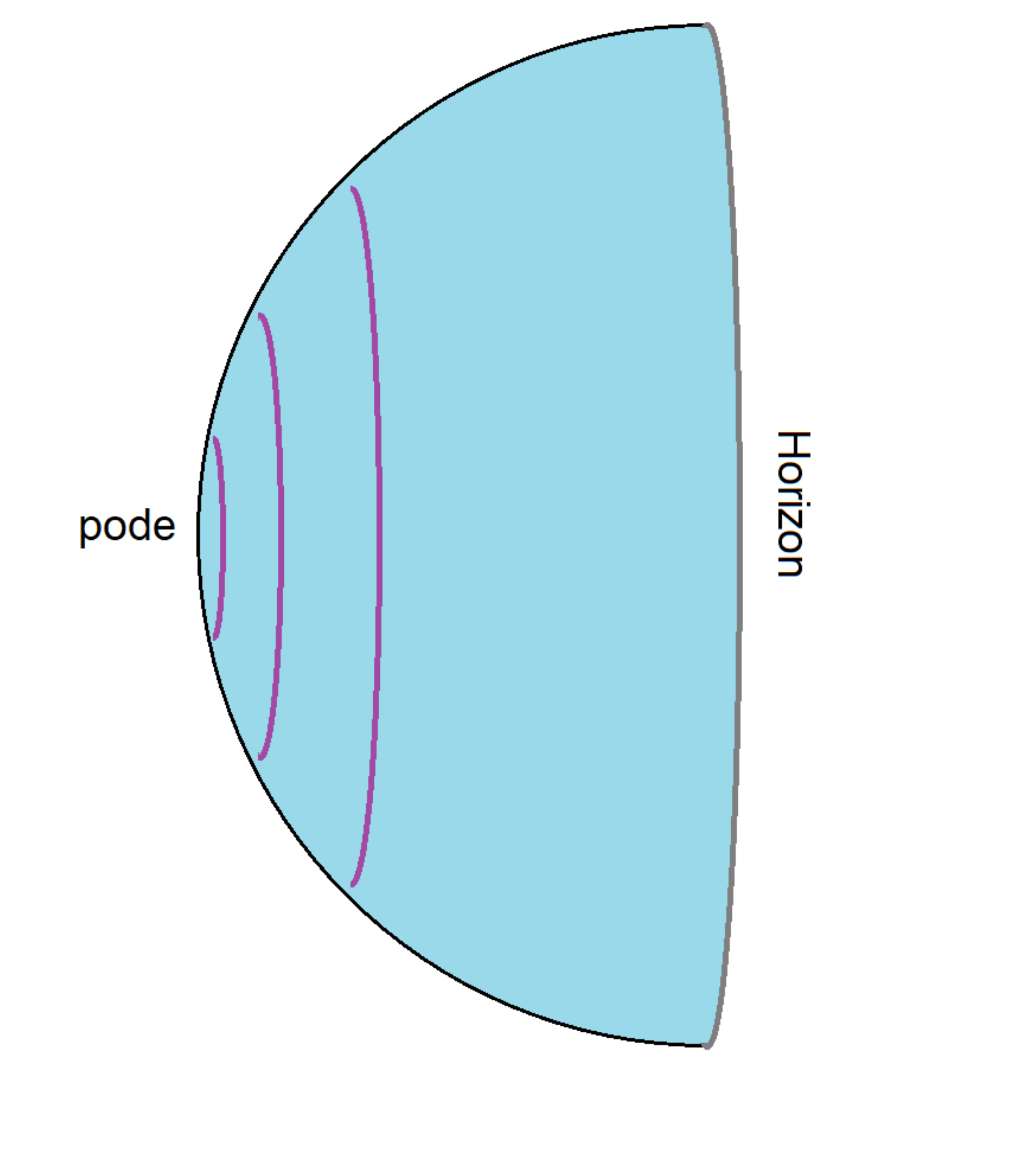}
\caption{Waves from the pode to the horizon }
\label{waves}
\end{center}
\end{figure}

 Now we may ask what do the oscillations look like to a causal patch observer? There are of course many causal patches but let's focus on the one indicated in blue in figure \ref{halfDS}.
 The answer is a very odd behavior illustrated in fig \ref{waves}. It is clear from figure
   \ref{halfDS} that the observer in the causal patch sees an oscillating behavior in the metric.
    The oscillations spread out from the
   pode\footnote{The pode refers to the point at the center of the causal patch} 
     with superluminal velocity and eventually reach the horizon. The horizon itself has an oscillating area.

Besides being very bizarre this behavior violates the second law. The eternal oscillations constitute a perpetual motion machine of the third kind, whereas the second law requires  the oscillations to  eventually dissipate and the system to come to thermal equilibrium.  This behavior   clearly violates  the second law, unless, of course the cosmological version of the central dogma is wrong.

\section{The Farhi-Guth-Guven Process}

The next example  is the Farhi-Guth-Guven theory of universe creation in the laboratory \cite{Farhi:1989yr}  (see also \cite{Fischler:1990pk}). Because this example involves following a causal patch as it falls behind a black hole horizon, there may be reasons to be  less certain about the application of the cosmological dogma, but I will assume it is valid

\begin{figure}[H]
\begin{center}
\includegraphics[scale=.3]{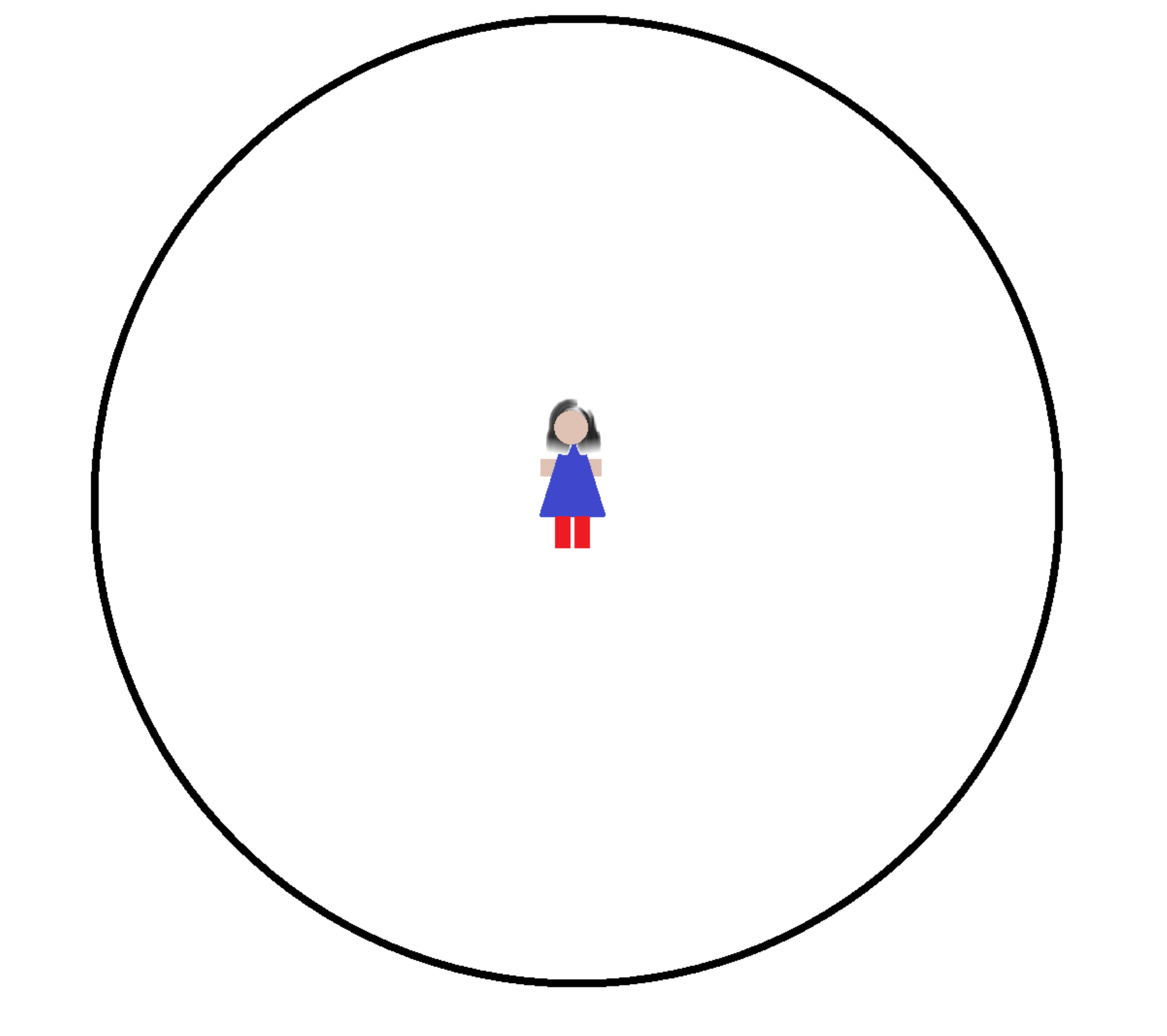}
\caption{Alice is initially at the center a large causal patch bounded by a very distant horizon}
\label{bigU}
\end{center}
\end{figure}

Without getting into the details of the Farhi-Guth-Guven process I will describe how an observer (Alice) might experience it. Let us suppose that initially Alice lives in a de Sitter space with a very small cosmological constant (fig \ref{bigU}), for example one similar to our own (Call this the initial state $A$). In that case the observable universe would have an entropy of order $S_A=10^{120}.$
\bn

\begin{figure}[H]
\begin{center}
\includegraphics[scale=.3]{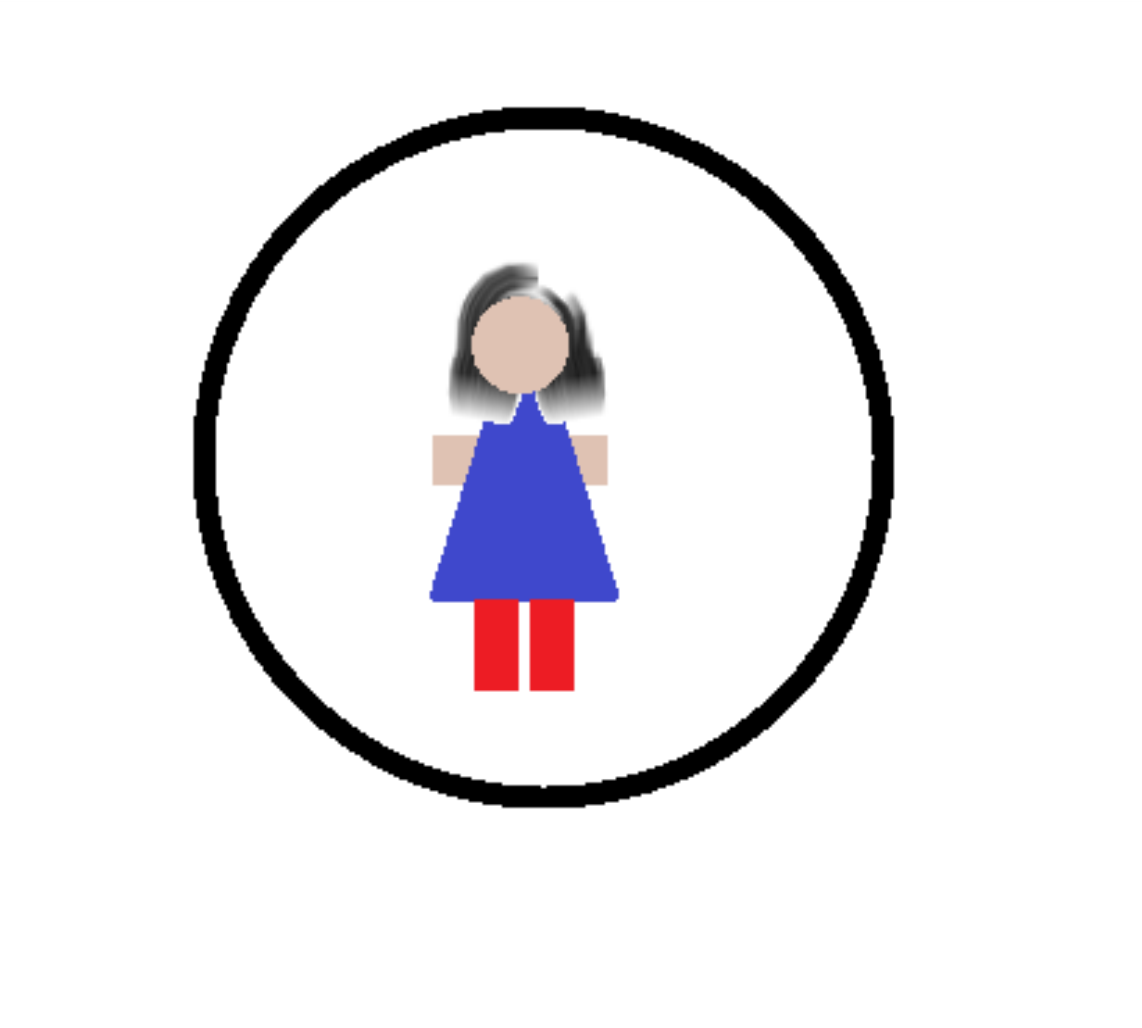}
\caption{After the Farhi-Guth-Guven process takes place Alice finds herself in a  smaller causal patch with a horizon of much smaller area.}
\label{smallU}
\end{center}
\end{figure}

Let's suppose Alice goes to sleep and while she is asleep Bob plays a trick on her.  He surrounds Alice with  a shell of matter. The shell is destined to form a black hole with Alice in the interior. According to \cite{Farhi:1989yr}\cite{Fischler:1990pk} a tunneling event may take place;   Alice, when she wakes,  finds herself in a de Sitter space with a much larger cosmological constant (final state $B$), and therefore a much smaller entropy than $S_A.$ Call it $S_B$ with $S_B<<S_A$ (fig \ref{smallU}).
\bn
Alice has witnessed a huge decrease in the entropy of her causal patch.

It was argued in \cite{Farhi:1989yr} the process is forbidden classically but can happen as a tunneling event. 
One may argue that such a tunneling event is a very rare fluctuation and rare fluctuations which decrease entropy are allowed. The problem is that according to \ref{P=expDS} the probability for witnessing such an event 
$$P_{A \to B} \approx e^{-(S_A-S_B)}$$
is vastly smaller than that predicted by \cite{Farhi:1989yr}\cite{Fischler:1990pk}. 
In fact \cite{Farhi:1989yr}\cite{Fischler:1990pk} claim a probability which is independent of $S_A,$ and  only depends on $S_B$ and the local properties of the matter which made up the in-falling shell. 

The Farhi-Guth-Guven theory, if correct, would lead to a very surprising conclusion for the long-time evolution of Alice's causal patch. First of all, it is not necessary for Bob to be involved at all. The black hole that engulfs Alice could be the result of a thermal fluctuation in the background de Sitter space $A$. The probability for such a fluctuation is,
$$ 
P_{bh} = e^{-M/T_A} = e^{-2\pi MR}.
$$
where $M$ is the black hole mass and $T_A$ is the temperature of the initial de Sitter space.
If we multiply this by the Farhi-Guth-Guven rate we find that FGG predict an ``up-tunneling" rate for the transition $A\to B,$
\be 
\bar{\Gamma}_{up}  = e^{-2\pi MR} \Gamma_{FGG}.
\ee 

On the other hand detailed balance requires the up-tunneling rate to be,
\be 
{\Gamma}_{up} = e^{S_B-S_A}\Gamma_{CDL}.
\ee
where $\Gamma_{CDL}$  is the Coleman-DeLuccia down-tunneling rate. The quantities $S_B,$ $M$, $\Gamma_{FGG}$, and $\Gamma_{CDL}$ are all independent of large-scale cosmology and have finite limits as $R\to \infty$. 
Only 
$S_A$ is sensitive to $R$. One finds that for very large $R$ the ratio of the FGG value of the up-tunneling rate, and the detailed balance value, is given by,
\be 
\frac{\bar{\Gamma}_{up} }{{\Gamma}_{up} } \approx \frac{\Gamma_{FGG}}{\Gamma_{CDL}S_B} e^{S_A} \sim e^{S_A}.
\ee
In other words the FGG rate is vastly too large to be consistent with detailed balance.

One must conclude that either there is something wrong, either with the analyses in  \cite{Farhi:1989yr}\cite{Fischler:1990pk},  or with the cosmological dogma.

\section{Remarks}

The central dogma of black hole physics  is widely accepted. It is very well supported by a long string of developments growing out of string theory, Matrix theory, and above all gauge-gravity duality. The central dogma of cosmological horizons is not nearly as well supported at this stage. By comparison with AdS space, the quantum mechanics of de Sitter space is poorly understood. Nevertheless the cosmological dogma  seems  plausible. If we accept it then the observations of an observer in a causal patch are constrained by the second law of thermodynamics. I've given arguments that three commonly studied theoretical ideas violate this principle.  

I should like to point out that the authors of both \cite{Farhi:1989yr} and \cite{Fischler:1990pk} discussed various reservations about their analyses,  although the second law was not one of them.

\section*{Acknowledgements}

I thank Alan Guth for discussions about the apparent conflict between the Farhi-Guth-Guven process and the second law. My argument given here, and some of our discussions, were described by Alan in his talk at 3rd Northeast String Cosmology
Workshop at Columbia University on May 14, 2004.

\end{document}